# Addressing Decadal Survey Science through Community Access to Highly Multiplexed Spectroscopy with BigBOSS on the KPNO Mayall Telescope

Caty Pilachowski (Indiana U), Carles Badenes (U of Pittsburgh), Stephen Bailey (LBNL), Aaron Barth (UC Irvine), Rachel Beaton (U of Virginia), Eric Bell (U of Michigan), Rebecca Bernstein (UC Santa Cruz), Fuyan Bian (U of Arizona), Michael Blanton (NYU), Robert Blum (NOAO), Adam Bolton (U Utah), Howard Bond (STScI), Mark Brodwin (U of Missouri), James Bullock (UC Irvine), Jeff Carlin (RPI), Ranga-Ram Chary (Caltech/IPAC), David Cinabro (Wayne State), Michael Cooper (UC Irvine), Jorge L. C. Cota (ININ, Mexico), Marc Davis (UC Berkeley), Kyle Dawson (U of Utah), Arjun Dey (NOAO), Megan Donahue (MSU), Jeremy Drake (CfA), Erica Ellingson (U Colorado), Lorenzo Faccioli (Kavli/Peking), Xiaohui Fan (U of Arizona), Harry Ferguson (STScI), Eric Gawiser (Rutgers), Marla Geha (Yale U), Mauro Giavalisco (U Mass), Anthony Gonzalez (U Florida), Kim Griest (UC San Diego), Bruce Grossan (LBNL), Raja Guhathakurta (UC Santa Cruz), Paul Harding (CWRU), Sara R. Heap (NASA/GSFC), Shirley Ho (Carnegie-Mellon), Steve Howell (NASA/Ames), Buell Jannuzi (Steward Observatory), Jason Kalirai (STScI), Brian Keeney (U Colorado), Lisa Kewley (U. Hawaii), Xu Kong (USTC), Michael Lampton (LBNL), Wei-Peng Lin (SAO), Axel de la Macorra (UNAM/IAC), Lucas Macri (Texas A&M), Steve Majewski (U Virginia), Paul Martini (OSU), Phil Massey (Lowell Obs), Virginia McSwain (Lehigh U), Adam A. Miller (UC Berkeley), Dante Minniti (P U Catolica), Maryam Modjaz (NYU), Heather Morrison (CWRU), John Moustakas (Siena College), Adam Myers (U of Wyoming), Joan Najita (NOAO), Jeffrey Newman (U Pittsburgh), Dara Norman (NOAO), Knut Olsen (NOAO), Michael Pierce (U Wyoming), Alexandra Pope (U Mass), Moire Prescott (UC Santa Barbara), Naveen Reddy (UC Riverside), Kevin Reil (SLAC), Armin Rest (STScI), Katherine Rhode (Indiana U), Connie Rockosi (UC Santa Cruz), Greg Rudnick (Kansas U), Abhijit Saha (NOAO), John Salzer (Indiana U), David Sanders (U Hawaii), David Schlegel (LBNL), Branimir Sesar (Caltech), Joseph Shields (Ohio U), Jeffrey Silverman (IPMU), Josh Simon (OCIW), Adam Stanford (UC Davis), Daniel Stern (JPL), Lisa Storrie-Lombardi (Caltech/IPAC), Nicholas Suntzeff (Texas A&M), Jason Surace (Caltech/IPAC), Alex Szalay (JHU), Melville Ulmer (Northwestern), Ben Weiner (U Arizona), Beth Willman (Haverford), Rogier Windhorst (ASU), Michael Wood-Vasey (U Pittsburgh)

**Executive Summary**: This document summarizes the results of a community-based discussion of the potential science impact of the Mayall+BigBOSS highly multiplexed multi-object spectroscopic capability. The KPNO Mayall 4m telescope equipped with the DOE- and internationally-funded BigBOSS spectrograph offers one of the most cost-efficient ways of accomplishing many of the pressing scientific goals identified for this decade by the "New Worlds, New Horizons" report. The BigBOSS Key Project will place unprecedented constraints on cosmological parameters related to the expansion history of the universe. With the addition of an open (publicly funded) community access component, the scientific impact of BigBOSS can be extended to many important astrophysical questions related to the origin and evolution of galaxies, stars, and the IGM. Massive spectroscopy is the critical missing ingredient in numerous ongoing and planned ground- and space-based surveys, and BigBOSS is unique in its ability to provide this to the US community. BigBOSS data from community-led projects will play a vital role in the education and training of students and in maintaining US leadership in these fields of astrophysics. *We urge the NSF-AST division to support community science with the BigBOSS multi-object spectrograph through the period of the BigBOSS survey in order to ensure public access to the extraordinary spectroscopic capability.*

## Introduction

In 2009 NOAO issued a Large Science Program Call[1] for the 4m Mayall telescope at Kitt Peak. This call resulted in the selection of BigBOSS[2], a project that aims to build (by 2018) a 5000-fiber, 3° diameter field of view, moderate resolution (R~3000-4800) multi-object spectrograph for the Mayall. The primary scientific goal of the BigBOSS collaboration is to conduct a Stage IV Dark Energy project by (a) measuring the baryon acoustic oscillation signature in galaxies and the Lyman alpha forest and (b) the growth of structure through an unprecedented redshift survey of more than 20 million galaxies and 600,000 quasars over 14,000 deg$^2$.

The Lawrence Berkeley National Laboratory leads an international collaboration that currently includes 34 US and foreign universities and laboratories to construct the BigBOSS instrument. The proposed BigBOSS Key Project will utilize approximately 500 nights on the Mayall telescope, distributed over 5 years. *The bulk of costs of construction and Key Project survey operations will be supported by non-NSF/AST funds*. As specified in the Large Science Program call, public access to BigBOSS will be available, contingent on continued support of the Mayall by NSF.

NSF investment in the Mayall is therefore a very cost-efficient way to provide to the US community an unprecedented capability that is highly relevant to the science goals described in the Astro2010's report "New

---

[1] See http://www.noao.edu/kpno/largescience.html
[2] See http://bigboss.lbl.gov/ and Schlegel et al. 2011, arXiv:1106.1706

Worlds, New Horizons". Both the archival Key Project data and the opportunities for PI-driven, community-led science programs enable the broad participation that drives scientific discovery and maintains the health of the astronomical profession.

In this submission, we describe the opportunities to address Decadal Survey science priorities through community-led, BigBOSS science programs. The science cases described herein were developed at a community workshop hosted by NOAO in 2011 that attracted approximately 70 participants. Attendees considered all modes of BigBOSS operations: the use of archival Key Project data, "synchronous" observing modes (i.e., using available "community" fibers during the BigBOSS Survey observations), and independent, PI-driven surveys with the instrument. Other details regarding the meeting, the presentation slides, and supporting documentation on BigBOSS can be found on the web[3].

The recommendation of the Portfolio Review to divest from telescopes on Kitt Peak places these opportunities at risk. The current document presents a brief recounting of only a few of the many extraordinary science opportunities enabled by BigBOSS and a plea to the NSF to enable these in a cost-efficient way by preserving public access to the BigBOSS instrument during the operational phase.

## BigBOSS and the Scientific Priorities for this Decade

The Astro2010 Decadal Survey report ("New Worlds, New Horizons" = NWNH) presents science questions in four broad themes: Discovery, Origins, Understanding Cosmic Order, and Frontiers of Knowledge. In what follows, we discuss the impact of BigBOSS on the first three science questions within each of these themes (the last theme is addressed by the BigBOSS Key Project Science). In particular, we briefly discuss a few examples of community science projects that, using various BigBOSS observing modes, can address these science themes.

### 1. Opportunities for New Discovery with BigBoss

The Decadal Survey highlights time domain astronomy as an area with great potential for new discoveries. While there are many ongoing and planned surveys of the time-variable universe, the need for spectroscopic follow-up is both acute and increasing. Access to community fibers on BigBOSS can provide the critical capability in support of time-domain astronomy, playing both discovery and supporting roles in the era of all-sky and synoptic imaging surveys.

For example, the BigBOSS Key Project survey data can, if cadenced appropriately, be used to identify rare spectroscopic variables such as short-period, white-dwarf binaries. White dwarfs are the main contaminant in the QSO candidate samples, and short-period white dwarf binaries can be discovered in archival data by identifying velocity shifts in consecutive 10-minute observations. Systems like the recently-discovered spectacular WD-WD binary with a 12-minute period[4] could be easily detected by BigBOSS and provide critical tests of gravitational wave astronomy and general relativity.

With rapid response modes integrated into Survey operations, community fibers could be allocated to observe time-critical transient events such as Advanced LIGO transients, ICECUBE transients, Gamma Ray Bursts (GRBs), or transients from the next generation of optical synoptic surveys. BigBOSS is unique in its capability to search for an electromagnetic counterpart to a gravitational wave signal. Advanced LIGO signals will be poorly localized on the sky (median of 10 square degrees) and the electromagnetic counterpart is expected to be short-lived. Therefore, spectra of hundreds of transient candidates that are spatially and temporally coincident will need to be taken within a few hours to unambiguously identify the true counterpart .

---

[3] http://www.noao.edu/meetings/bigboss/
[4] Brown et al. 2011, ApJ, 737, L23; see also Roelofs et al. 2010, ApJ, 711, L138; Kilic et al. 2010, ApJ, 716, 122

Similarly, ICECUBE TeV neutrino signals are also localized to a couple of square degrees. The field of view and sensitivity of BigBOSS are ideally suited to this search. GRBs are linked to the most distant objects and the most energetic transient events known, possibly representing extremes of particle acceleration and production. Despite their importance, GRB follow-up (especially short GRBs) is starved for spectra, with the percentage of GRBs with spectra small and dropping. Many GRBs are brighter than 21.1 mag (visual) and thus easily observed with BigBOSS, if the timing is consistent with the survey strategy.

Alternatively, an imaging transient survey for a variety of transients could precede the BigBOSS survey observations if the observing strategy and cadence are known sufficiently in advance. This approach could target a variety of transient phenomena (novae, CVs, SN, etc.), perhaps with different priorities depending on their timescales, visibility, etc. Based on the SDSS-SN experience, imaging 150 deg$^2$ per night yielded ~100 transients. Since BigBOSS would observe ~1/2 the area of SDSS-SN to a deeper depth, there would be ~100 transient phenomena observed each night at essentially no cost to the Key Project survey.

Finally, the BigBOSS archive can be used in support of other time-domain studies, e.g., providing redshifts for host galaxies of SNe identified in photometric surveys and completing the census of galaxies in the local universe.

## 2. The Origin and Evolution of Astronomical Objects

This second theme highlighted in the Decadal Survey is an area where BigBOSS can certainly play key roles, e.g., in understanding star formation, the origin and evolution of the Milky Way and M31, and the origin and evolution of galaxies and galaxy clusters.

**Origin and Evolution of Stars** - Massive O- and B-type stars, the hottest and most massive classes of stars, drive much of the energy and chemical feedback into the Universe, a fundamental process identified in NWNH. Spectral classification of massive stars is critical for studies of the initial mass function because of degeneracy in the colors. BigBOSS would allow for unparalleled study of the populations of emission-line stars not only in the Milky Way, but also in M31 and M33, including Wolf-Rayet stars, B[e] stars, Luminous Blue Variable candidates, and many fainter evolved stars. BigBOSS will enable unprecedented survey of the rotation and variability of mass loss/winds from massive stars. These various disk and inner halo targets could be easily included as part of the survey of the entire M31 halo. M31 also includes thousands of catalogued HII regions associated with star formation that would allow for detailed kinematic and gas-phase abundance studies of the disk and inner halos of these neighbor galaxies. Strong-line abundances for HII regions are easily obtained, but pushing deep could greatly increase the number of HII regions with measured electron temperatures across the disk, leading to significantly more robust abundances. BigBOSS also enables the search for extremely rare stars in the Milky Way halo such as high velocity or extremely metal poor stars that cannot be identified via broad-band colors alone.

NWNH also identifies rotation in stars as a key unsolved problem. How does fragmentation play a role in the distribution of mass and angular momentum during the star formation process? How does angular momentum evolve during the main sequence and later evolution? Rich fields such as Cyg OB2 that are the current targets of binarity and angular momentum surveys are well-matched to the BigBOSS fiber spacing and large field of view. The BigBOSS spectrographs can resolve velocities of 75-100 km s$^{-1}$ to study the angular momentum of hot stars (with $<v\sin i> \sim 200$ km s$^{-1}$). For a well-calibrated, stable instrument, recent studies have shown that order of magnitude improvements on this precision are possible. By measuring the mass distribution, multiplicity fraction, and angular momentum distribution of large numbers of cluster members, BigBOSS can provide a critical test of the role of angular momentum in star formation and binary fragmentation theory.

**Origin and Structure of Nearby Galaxies** - Studies of the Milky Way and nearby resolved galaxies such as M31 and M33 contribute to NWNH goals to understand the origin and evolution of galaxies as well as "near-field" cosmology--the examination of the Milky Way and Local Group to understand the formation and assembly of galaxies, chemical evolution, and dark matter distribution. In contrast to most extragalactic studies based on integrated light, in Local Group galaxies one can observe individual stars. This enables studies to far deeper effective surface brightness levels and more accurately isolates various galactic components (e.g., the disk, bulge, halo or individual tidal streams). Studies based on individual stars require large samples over wide regions of the sky in order to map the kinematic and chemical distributions of various galactic subcomponents.

BigBOSS's wide field-of-view, large wavelength coverage, and highly multiplexed capabilities make it a transformational capability in the study of near-field cosmology and galaxy assembly. BigBOSS surveys of the MW halo can discover and characterize the tenuous, low-surface brightness tidal streams, and thereby reconstruct the assembly history of the halo. Similarly, BigBOSS surveys of the inner Galactic thick disk and bulge offer an order-of-magnitude improvement over any other planned survey, and can provide unprecedented constraints on the age distributions, age-metallicity relations, and phase space structure of these components.

Our vantage point offers a birds-eye view of M31, our nearest large spiral neighbor, to investigate spiral galaxy formation. M31 is far enough away that its entire structure is revealed, yet close enough that individual stars can be studied as fossils of the galaxy formation process. A key science goal is to understand the formation of M31's halo by characterizing substructure through the identification and measurement of the kinematics of streams, extending beyond 100 kpc from the nucleus. This program can also constrain the mass and mass distribution of M31's dark halo. Existing spectroscopic coverage of M31, limited to deep pencil beams through the M31 disk and halo, have provided a tantalizing look at the complex kinematic and chemical structure of M31. A *complete* map of the kinematics and abundances of the M31 halo, only feasible with BigBOSS, would be a tremendous step forward.

**Origin and Evolution of Galaxies** – NWNH emphasizes the importance of large spectroscopic surveys in the visible and near-infrared to trace the evolution of the distribution of galaxies and their dark matter halos, to follow star formation over cosmic time and in extreme environments, and to investigate chemical enrichment through the history of the Universe.

BigBOSS can contribute to all of these efforts in unparalleled ways. While the BigBOSS Key Project will yield an unprecedented galaxy redshift catalog, many galaxy evolution studies will require higher signal-to-noise ratio spectra and better sampling (for environmental studies). By complementing the BigBOSS Key Project calibration field observations with deeper/denser spectroscopic surveys, the community can investigate the growth of galaxy populations within the dark matter halos. Sophisticated algorithms have been developed to analyze large spectroscopic catalogs and probabilistically assign galaxies to dark matter halo masses (or mass ranges) and to identify their role within the halo (i.e., as satellite or central objects). Hence, BigBOSS surveys will result in empirical constraints on how "gastrophysical" processes (e.g., star formation, AGN activity) and evolutionary phases (e.g., post-starburst) depend on the (sub-)halo mass and epoch. The same survey could result in large samples of rare objects (e.g., bright Lyman break galaxies, IR-luminous galaxies, very luminous starbursts, QSOs, etc.), that offer the ability for detailed study and thus valuable insights into the physical processes driving galaxy evolution. Such rare objects might be identified in imaging surveys and included (for "synchronous mode" observations) as part of BigBOSS survey observations.

**Galaxy Cluster Evolution -** New generation cluster catalogs are rapidly becoming available via several techniques, including optical, IR, Sunyaev-Zeldovich and X-ray selection. While photometric redshifts from the red sequence in cluster cores can provide cluster redshifts to few-percent accuracy, a single BigBOSS fiber placed on brightest cluster galaxy will increase this accuracy significantly, leveraging multi-wavelength observations to determine cluster properties. Additional fibers targeting luminous red galaxies (LRGs) can identify associated cluster substructure within a wide field of view. At $0.05 < z < 0.2$, the BigBOSS field of view is an excellent match to the virialized regions of massive galaxy clusters. A deep redshift survey of >1000 galaxies to several virial radii will map cluster caustics and measure in detail the cluster dark matter mass profile to large radii. These profiles are crucial in testing formation and growth scenarios for dark matter halos and mapping the dark matter and baryon distributions on the largest scales. At $0.2 < z < 0.4$, BigBOSS maps both the virial and infall regions of clusters, allowing an unprecedented view of the galaxies which are the precursors of the populations that now inhabit the cores of massive clusters. Spectroscopic redshifts will allow identification of infalling groups and filaments to tens of Mpcs from cluster cores and will quantify the effects of both local and global environments on galaxy properties. While SDSS continues to map structures on these largest scales, denser observations will be needed in the cluster fields to unambiguously identify and estimate the masses of dense infalling groups and subclusters.

### 3. Understanding the Cosmic Order

The majority of baryons in our universe (by volume and mass) exist within a diffuse medium outside stars and galaxies. Investigations of the chemistry, kinematics, and structure of the diffuse media bear directly on the understanding of the flows of baryons into/out of galaxies, feedback mechanisms which are critical to studies of galaxy evolution, the chemical enrichment of galaxies (and the universe), and the mass/energy/chemical cycles within galaxies. The BigBOSS Key Project, which determines redshifts of more than 20 million galaxies *and* 600,000 background quasars, naturally provides an invaluable database for the study of the diffuse media over a large range in redshift ($0.3 < z < 5$).

Community fibers (used in "synchronous mode") on BigBOSS targeting many galaxies along/near the line of sight to known QSOs can provide the opportunity to explore outflows in $z\sim0.5$ star-forming galaxies, while simultaneously constraining the spatial extent of the circumgalactic medium (CGM) surrounding the galaxies. The (stacked or individual) galaxy spectra in combination with QSO absorption lines can provide a (statistical) map of the surrounding CGM. The same program could also yield close galaxy-QSO, QSO-QSO, or galaxy-galaxy pairs, which can provide unique data on the CGM on small scales.

Dedicated BigBOSS community surveys of large samples of background galaxies behind nearby galaxies (i.e., with optical disk radii of 5 to 10 arcmin, and therefore virial radii of $\sim$ 1 degree) can map the differential extinction and the gas and dust ejecta in the CGM. Finally, as exemplified by studies of the SDSS dataset, archival studies of the BigBOSS datasets will likely yield new insights on the dust extinction in the CGM and ISM of galaxies, discoveries of new gravitational lens candidates (for follow-up investigations of the ISM or mass determinations), and (through stacking analyses) novel constraints on the mean opacity of the IGM.

### Broader Impact

The Mayall telescope is a mainstay of US astronomy, supporting the research and educational efforts of a large, diverse community of US optical/infrared astronomers. It is one of the only 4m telescopes capable of a $3^o$ wide-field and supporting the BigBOSS instrument. BigBOSS is a powerful capability that is highly aligned with a many of the science goals of "New Worlds, New Horizons." Thus, BigBOSS provides an opportunity for NSF to support at minimal cost (i.e., the continued operation of the Mayall telescope) a world-beating capability for the US astronomical community that enables the participation of a broad community in the pursuit of some of the most pressing science goals of the coming decade.

While it might be possible for the BigBOSS Collaboration to obtain funding to undertake their Key Project on a dedicated Mayall telescope, the loss to the US community, to NWNH goals, and to science productivity would be grave. Most of the science cases described here (and at the Community Workshop) require access to the instrument for specific surveys, or access to "community fibers" during the BigBOSS Key Project observations. These would only be possible if public access to the project is maintained. *We strongly urge the NSF-AST Division to ensure public access to this powerful new capability for community science.*